\documentclass[11pt,a4paper]{article}
\usepackage{jheppub}
\usepackage{textcomp}

\newcommand{\lag}{\mathcal{L}}
\newcommand{\AFB}{A_{\text{FB}}}
\def\msbar{$\overline{\hbox{MS}}$}

\usepackage{multirow}
\usepackage{caption}
\usepackage{subcaption}
\usepackage{placeins}

\newcommand{\seff}{\sin \theta_\text{eff}^\ell}

\def\msbar{$\overline{\hbox{MS}}$}

\title{Analytic results for electroweak precision observables at NLO in SMEFT}
\preprint{IPPP/25/13, KIT-TP-04-2025, P3H-25-015}

\author[a]{Anke Biek\"otter,}
\author[b]{Benjamin D.~Pecjak}
\affiliation[a]{Institute for Theoretical Physics, Karlsruhe Institute of Technology, 76131 Karlsruhe, Germany}
\affiliation[b]{Institute for Particle Physics Phenomenology, 
Durham University,  Durham DH1 3LE, UK}

\abstract{We present analytic results for electroweak precision observables (EWPO) at next-to-leading order (NLO) in dimension-six SMEFT, with no assumptions on the flavour structure of 
SMEFT Wilson coefficients. The results are given in five different electroweak input schemes, thus offering a simple means, along with scale variations, of estimating theory uncertainties related to higher-order terms in the SMEFT expansion. Our results will be useful to assess the constraining power of existing and future lepton colliders  for new physics scenarios.
}

\begin{document}
\maketitle

\section{Introduction}
\label{sec:intro}

Standard Model Effective Field Theory (SMEFT) is a robust and widely used framework for describing small deviations from the predictions of the Standard Model (SM) of particle physics~\cite{Brivio:2017vri}. To increase the precision of SMEFT predictions, higher-order corrections in the SM couplings or, equivalently, loops need to be included. 
The calculation of next-to-leading order~(NLO) SMEFT corrections is an active area of research: NLO QCD corrections have been fully automated for most Wilson coefficients~\cite{Degrande:2020evl,Rossia:2024rfo}
and NLO electroweak as well as NNLO QCD corrections have been calculated for various processes on a case-by-case basis in~\cite{Zhang:2013xya,Crivellin:2013hpa,Zhang:2014rja,Pruna:2014asa,Grober:2015cwa,Hartmann:2015oia,Ghezzi:2015vva,Hartmann:2015aia,Aebischer:2015fzz,Zhang:2016omx,BessidskaiaBylund:2016jvp,Maltoni:2016yxb,Degrande:2016dqg,Hartmann:2016pil,Grazzini:2016paz,deFlorian:2017qfk,Deutschmann:2017qum,Baglio:2017bfe,Dawson:2018pyl,Degrande:2018fog,Vryonidou:2018eyv,Dedes:2018seb,Grazzini:2018eyk,Dawson:2018liq,Dawson:2018jlg,Dawson:2018dxp,Neumann:2019kvk,Dedes:2019bew,Boughezal:2019xpp,Dawson:2019clf,Baglio:2019uty,Haisch:2020ahr,David:2020pzt,Dittmaier:2021fls,Dawson:2021ofa,Boughezal:2021tih,Battaglia:2021nys,Kley:2021yhn,Faham:2021zet,Haisch:2022nwz,Heinrich:2022idm,Bhardwaj:2022qtk,Asteriadis:2022ras,Bellafronte:2023amz,Kidonakis:2023htm,Gauld:2023gtb,Heinrich:2023rsd,Asteriadis:2024xuk,Asteriadis:2024xts,Dawson:2024pft,ElFaham:2024egs}. 
As emphasised in \cite{Biekotter:2023xle}, the
perturbative convergence and pattern of SMEFT Wilson coefficients appearing in  higher-order corrections is influenced by the choice of electroweak input scheme, so evaluating 
observables in different schemes gives a means of assessing theoretical uncertainties 
beyond scale variations alone.  

In this paper, we present analytic NLO results in dimension-6 SMEFT for the so-called electroweak precision observables~(EWPO) measured at LEP and the Tevatron~\cite{ALEPH:2005ab}.  In particular, we extend the NLO SMEFT predictions presented in~\cite{Dawson:2019clf,Dawson:2022bxd,Bellafronte:2023amz} in a (mostly) numerical form in
the $\{G_F , \alpha(0) , M_Z \}$ input scheme to five different electroweak schemes 
involving $M_Z$ and combinations of $G_F$, $M_W$, $\alpha(M_Z)$ and $\seff$ as inputs, in
each case providing fully analytic results with no flavour assumptions on the SMEFT Wilson coefficients. Given that the EWPO currently provide some of the most precise probes of new physics and are expected to be measured with significantly increased precision at a future $e^+ e^-$ collider like FCC-ee or CEPC~\cite{FCC:2018evy, CEPCStudyGroup:2018ghi},  these results
will be useful in assessing theory uncertainties and providing cross-checks on near-term and future
global SMEFT fits.\footnote{Indeed, they have already been used in the recent analysis of~\cite{Mildner:2024wbl}.}  While the methods used in obtaining our results follow previous work~\cite{Biekotter:2023xle,Biekotter:2023vbh}, where the leptonic decay rates of the 
$Z$ and $W$ bosons were calculated, we have extended those calculations
by computing decays into hadrons and neutrinos and taking into account the chiral
structure needed to obtain left-right and forward-backward asymmetries. 

The paper is structured as follows.  First, in section~\ref{sec:details}, we give some calculational
details, defining NLO SMEFT expansions of the EWPO and the different electroweak input schemes in which the calculations are performed.  In section~\ref{sec:results} we present our results, explaining
the notation and contents of ancillary electronic files as well as how to evaluate them numerically
for arbitrary input parameters, including uncertainty estimates.  We conclude in Section~\ref{sec:conclusions} and provide in appendices further details of our calculations: in appendix~\ref{sec:EWPO_def} we define the EWPO on the $Z$-pole, appendix~\ref{sec:flav} covers
different flavour assumptions provided along with our results, appendix~\ref{sec:comparison_prev} compares with previous literature, and in 
appendix~\ref{sec:EWPO_LO} we give simple analytic expressions needed to evaluate
 EWPO at LO in  SMEFT in the five different input schemes. The analytical results at NLO calculated in this work are provided as ancillary files with the arXiv submission.

\section{Calculational Details}
\label{sec:details}

We write the dimension-six SMEFT Lagrangian as 
\begin{align}
\lag = \lag^{(4)} + \lag^{(6)}  ;  \quad \lag^{(6)} =  \sum_i C_i(\mu) \, Q_i(\mu) \, ,
\end{align} 
where $ \lag^{(4)}$ denotes the SM Lagrangian and $ \lag^{(6)}$ is the 
dimension-six Lagrangian with operators $Q_i$ given in the Warsaw basis~\cite{Grzadkowski:2010es} and the corresponding 
Wilson coefficients $C_i(\mu)\equiv C_i = c_i/\Lambda^2$ 
are inherently suppressed by the new  physics scale $\Lambda$.  
The 59 independent dimension-six operators, which in general carry flavour indices, are listed and grouped into eight classes in Table~\ref{op59}.\footnote{We employ the symmetric basis for the Wilson coefficients. This means that for Wilson coefficients contributing to operators with two identical fermion bilinears, we define the Wilson coefficient of both flavour combinations and take into account their symmetry, e.g.\ $C_{\substack{ll\\1221}} + C_{\substack{ll\\2112}} = 2 \, C_{\substack{ll\\1221}} $.}
Throughout this work, we truncate the SMEFT expansion of a given quantity to linear order in the Wilson coefficients. 

We assume that the CKM matrix is the unit matrix and that all fermions are massless except the top quark with mass $m_t$. Given the expansion to linear order in the Wilson coefficients, no flavour-violating SMEFT interactions contribute, as we rely on the interference with the corresponding SM diagrams.  We make no assumptions on the flavour structure of the SMEFT interactions, but we provide replacement rules to obtain the results for a $U(3)^5$ symmetry of the fermion fields as well as in minimal flavour violation in appendix~\ref{sec:flav}.
 
\begin{table}
	\centering
	\begin{tabular}{c | l l }
	 scheme $s$ & inputs & suffix in filenames \\ \hline
		 $v_\mu^{\rm eff}$ & $G_F$, $\seff$, $M_Z$ & \texttt{GF\_SWeff\_MZ}  \\ 
		
		$v_\alpha^{\rm eff}$ &  $\alpha(M_Z)$, $\seff$, $M_Z$ & \texttt{aEW\_SWeff\_MZ}\\ 
		
		 $\alpha_\mu$ &  $G_F$, $M_W$, $M_Z$ & \texttt{GF\_MW\_MZ}\\
		
		 $\alpha$ &  $\alpha(M_Z)$, $M_W$, $M_Z$& \texttt{aEW\_MW\_MZ}\\
		
		 LEP &  $G_F$, $\alpha(M_Z)$, $M_Z$ & \texttt{GF\_aEW\_MZ}
		 
	\end{tabular}
	\caption{\label{tab:schemeDef} Nomenclature for the EW input schemes considered in this work.}
\end{table}
We consider the five different electroweak input schemes listed in Tab.~\ref{tab:schemeDef}, 
which use as inputs a combination of three parameters in the following list: the Fermi constant~$G_F$, the masses of the $Z$ and $W$ bosons $M_Z, \, M_W$, the leptonic effective mixing angle $\seff$ and  the electromagnetic coupling constant, for which we use the
on-shell definition $\alpha(M_Z)$. This definition is closely related to the \msbar~definition
$\bar{\alpha}(\mu)$ in five-flavour QED$\times$QCD used in~\cite{Biekotter:2023xle}, where the electroweak scale contributions are included through decoupling constants as described in Section 4.2 of \cite{Cullen:2019nnr}.  Explicitly, the definitions are linked via the perturbative relation
\begin{align}
\label{eq:Alpha_Convert}
\overline{\alpha}(\mu) &  = \alpha(M_Z)\left[1+ \frac{\alpha(M_Z)}{\pi} \sum_{f\neq t} \frac{N_c^f}{3}Q_f^2\left(\frac{5}{3} +  \ln \frac{\mu^2}{M_Z^2}\right) \right]  \nonumber \\
 & =\alpha(M_Z)\left[1+ \frac{\alpha(M_Z)}{\pi} \left(\frac{100}{27} +  \frac{20}{9}\ln \frac{\mu^2}{M_Z^2}\right) \right]   \, ,
\end{align}
where $Q_f$ is the charge of the fermion and $N_c^f=3$ ($N_c^f =1$)  for quarks (leptons).  
In practice, the numerical value of $\alpha(M_Z)$ can either be taken from a fit, or else calculated from $\alpha(0)\approx 1/137$ through the relation
\begin{align}
\label{eq:alpha_relations}
\alpha (M_Z) &= \frac{\alpha(0)}{1- \Delta \alpha(M_Z)} \, ,
\end{align}
with $\Delta \alpha (M_Z) = \Delta \alpha_\text{lep}+ \Delta \alpha_\text{had}= 0.03142 + 0.02783$,
where the leptonic contribution has been obtained by evaluating the one-loop 
contribution
\begin{equation}
\Delta \alpha_\text{lep} = \frac{\alpha}{3\pi}\left( - \frac{15}{3} + \sum_{\ell = e, \mu, \tau} \log \frac{M_Z^2}{m_\ell^2} \right) \, ,
\end{equation}
and the value of the hadronic contribution is taken from~\cite{ParticleDataGroup:2024cfk}. In this latter case,
one can view $\alpha(0)$ as the fundamental input parameter, while the use of 
$\alpha (M_Z)$ affects a resummation of light-fermion contributions which circumvents their
explicit appearance in the fixed-order SMEFT expansion coefficients. Our use of $\alpha(M_Z)$
rather an $\alpha(0)$ as an input parameter in the LEP scheme is one of the differences
compared to the  corresponding NLO SMEFT calculations of EWPO in \cite{Dawson:2019clf,Dawson:2022bxd,Bellafronte:2023amz}; other differences are explained
in Appendix~\ref{sec:comparison_prev}.

To shorten analytic expressions and make clear the origin of certain terms, we make use of dependent parameters whose explicit expressions are input-scheme
dependent.  For instance, the sine and cosine of the Weinberg angle 
appearing in our equations depend on the electroweak input scheme $s$, as does the 
$W$-boson mass, according to the following:
\begin{align}
\label{eq:CW_def}
\left(c_w^s\right)^2 = 1-\left(s_w^s\right)^2 = \frac{\left(M_W^s\right)^2}{M_Z^2} =
\begin{cases}
\frac{M_W^2}{M_Z^2} \, ,& s=\{\alpha,\alpha_\mu\} \\
1-\left( \seff \right)^2 \, ,&  s=\{v_\alpha^{\rm eff},v_\mu^{\rm eff}\} \\
\frac{1}{2} \left( 1 + \sqrt{1- \frac{4 \pi \alpha(M_Z) v_\mu^2}{M_Z^2}} \right) \, , & s={\rm LEP}  
\end{cases}
\,.
\end{align}
The LEP-scheme result in Eq.~(\ref{eq:CW_def}) depends on the quantity $v_\mu$,
which is related to the vacuum expectation value (vev) of the Higgs field at LO in the SM. This
follows the notation of \cite{Biekotter:2023xle,Biekotter:2023vbh}, which denotes the vev in scheme $s$ by $v_s$, where in this case the explicit expressions are
\begin{align}
\label{eq:vs_def}
v_{\alpha_\mu} &= v_{v_\mu^{\rm eff}} = v_{\rm LEP} \equiv v_\mu = \left( \sqrt{2}G_F \right)^{-\frac12} \, ,  \nonumber \\
v_\alpha & =\frac{ 2M_W s_w}{\sqrt{4\pi \alpha(M_Z)}} \, , \quad v_{v_\alpha^{\rm eff}} = \frac{2 M_Z c_w s_w}{\sqrt{4\pi \alpha(M_Z)}} \,.
\end{align}
Note that we have dropped the superscript $s$ in factors of $c_w$ and $s_w$ appearing in the $v_\alpha^{\rm eff}$-scheme result, which must be understood in the sense of Eq.~(\ref{eq:CW_def}).
Given that it is always clear from the context which scheme $s$ is under consideration, we follow this convention in the remainder of the paper and the associated electronic files.

\subsection{Observables and the NLO SMEFT expansion}
The considered EWPO are based on partial $Z$ and $W$ boson decay rates. 
For the $W$ boson, we provide its total decay rate, as well as the partial lepton decay rates
\begin{align}
 \Gamma_W ,  \, \Gamma_W^{\ell\nu}  \, ,
\label{eq:EWPO_W}
\end{align}
with $\ell = e, \, \mu, \, \tau$. 
On the $Z$ pole, we consider the total $Z$-boson decay rate, the hadronic total cross section in the narrow width approximation, as well as ratios, left-right asymmetries and forward-backward asymmetries for decays into leptons, bottom, charm, and strange quarks
\begin{align}
\Gamma_Z, \, \sigma_{\text{had}} , \, 
R_\ell , \, R_q  , \, 
A_\ell, \,  A_q, \, 
\AFB^\ell , \,\AFB^q 
\, ,
\label{eq:EWPO_Z}
\end{align}
with $q=s,c,b$.
The definitions of these quantities are given in appendix~\ref{sec:EWPO_def}.   
In addition, we present predictions for 
\begin{align}
M_W, \, G_F , \, \alpha 
\label{eq:EWPO_add}
\end{align}
in those schemes in which they are not an input parameter.  Note that $\seff$ is not included
in the list Eq.~(\ref{eq:EWPO_add}), because as shown in Eq.~(\ref{eq:seff_Aell}) it is 
equivalent to $A_{\ell}$. For this reason, in input schemes involving $\seff$, one 
must exclude $A_{\ell=L_{\rm eff}}$ from the list of independent observables, where
$L_{\rm eff}$ is the reference lepton used in defining the effective Weinberg angle.

We expand all observables to linear/first order in the EFT and loop expansion and consistently
drop all partial higher-order corrections. For an observable $O$ we define LO and NLO
results in terms of SMEFT expansion  coefficients  in scheme $s$ as 
\begin{align}
\label{eq:SMEFT_expansion}
O_s^{\rm LO}&=  O_s^{(4,0)}+ v_s^2 \,   O_s^{(6,0)} \, , \nonumber \\
O_s^{\rm NLO}  &= O_s^{\rm LO} + \frac{1}{v_s^2} O_s^{(4,1)}  + O_s^{(6,1)} \, ,
\end{align} 
where the superscripts $l$ and $k$ in $O_s^{(l,k)}$ label the operator dimension and the number of loops ($k=0$ for tree-level and $k=1$ for one-loop), respectively.
The expansion above makes the dependence on the vev explicit, as electroweak loop contributions are proportional to two powers of the coupling $g^2 \sim 1/v_s^2$.

\subsection{Tools and conventions}
SMEFT Feynman rules have been obtained using an in-house \texttt{FeynRules}~\cite{Alloul:2013bka} implementation of the dimension-six SMEFT Lagrangian, and cross checked with \texttt{SMEFTsim}~\cite{Brivio:2017btx,Brivio:2020onw}. 
In contrast to our previous works~\cite{Biekotter:2023xle,Biekotter:2023vbh}, we define the covariant derivative with a plus sign to match the conventions of \texttt{SMEFTsim}~\cite{Brivio:2017btx,Brivio:2020onw}\footnote{This flips the sign of the predictions for the Wilson coefficients $C_W$ and those of the dipole operators.} so that, for instance,
\begin{align}
D_\mu q &= \left[ \partial_\mu + i g_s T^a G_\mu^a + i \frac{g_W}{2} \sigma^i W_\mu^i + i \mathbf{y}_q g_1 B_\mu \right] q  \, .
\end{align}
	Matrix elements were computed using~\texttt{FeynArts} and~\texttt{FormCalc}~\cite{Hahn:2016ebn,Hahn:1998yk,Hahn:2000kx} and analytic results for Feynman integrals were extracted from \texttt{PackageX}~\cite{Patel:2015tea}. We express the Passarino-Veltmann~(PV) integrals in the notation of \texttt{LoopTools}~\cite{Hahn:1998yk}.  In schemes in which the $W$-boson mass is not an input parameter, we have consistently expanded the $M_W$-dependent PV integrals for the SM one-loop contributions to linear order in the SMEFT. 
	Phase-space integrals arising from the real emission of photons and gluons were calculated analytically using standard methods. 

\subsection{Treatment of divergences and cross-checks}
UV and IR divergences appearing in the NLO calculation are treated in dimensional regularization in $d=4-2 \epsilon$ space-time dimensions, where $\epsilon$ is the dimensional regulator. 
IR divergences in the $\epsilon$ expansion cancel between virtual and real emission corrections, while  UV divergences are cancelled by adding appropriate counterterms. 
Tadpoles are treated in the FJ tadpole scheme~\cite{Fleischer:1980ub} and cancel 
out of results for physical observables. 
The SMEFT Wilson coefficients are renormalised in the \msbar~scheme, i.e.\ we relate the bare parameters, denoted with a subscript 0, to the renormalised quantities via the equation
\begin{align}
C_{i,0} & = C_i + \delta C_i , &  \delta C_i & \equiv \frac{1}{2 \epsilon} \dot{C}_i \equiv \frac{1}{2 \epsilon} \frac{\text{d} C_i}{\text{d} \log \mu} \, .
\end{align}
The $\dot{C}_i$ have been calculated at one loop in \cite{Jenkins:2013zja,Jenkins:2013wua,Alonso:2013hga}\footnote{We use the electronic implementation in \texttt{DsixTools}~\cite{Celis:2017hod, Fuentes-Martin:2020zaz} as the $\dot{C}_i$ typically depend on a large number of Wilson coefficients.}
and when written in terms of mass-basis parameters take the form
\begin{align}
\label{eq:Cdot_gamma}
\dot{C}_i = \frac{1}{v_s^2}\gamma_{ij}^{(4,1)} C_j \,,
\end{align}
which makes clear that changing $\mu$ mixes the Wilson coefficients in a way determined
by the one-loop anomalous dimension matrix $\gamma_{ij}^{(4,1)}$.

The masses $M_W$ and $M_Z$ are renormalised on-shell, relating the renormalised parameters to the bare ones according to 
\begin{align}
M_V^2 &= M_{0,V}^2 - \Pi_{VV} (M_V^2) \, ,
\end{align}
where $\Pi_{VV}(M_V^2)$ is the one-loop two-point function for the $Z$ or $W$ boson.
$G_F$ is renormalised by requiring that the relation $v_\mu = (\sqrt{2} G_F)^{-1/2} $ holds to all orders in perturbation theory, see appendix~A of~\cite{Biekotter:2023xle}. 
For the effective leptonic mixing angle, the renormalisation condition is $\seff = s_w$ and the necessary counterterms are listed explicitly in~\cite{Biekotter:2023vbh}. Finally, counterterms 
from electric charge renormalisation are obtained by renormalising the photon-fermion
vertex as described in \cite{Cullen:2019nnr}.

As internal cross-checks on our calculation we have verified the cancellation of all UV and IR 
divergences and arrived at identical results in unitary and Feynman gauge. We have also checked
that  the $\mu$-dependence in SMEFT corrections is as dictated by a renormalisation-group
analysis. In particular, while the LO expansion coefficients of a given observable 
contain only implicit $\mu$-dependence through the Wilson coefficients, 
$O_s^{(6,0)}= O_s^{(6,0)}(C_i(\mu))$, the NLO corrections also contain explicit 
$\mu$-dependence through the UV-renormalised matrix elements. The fact that 
the NLO decay rate is independent of $\mu$ to that order then requires that the NLO 
coefficients take the form
\begin{align}
\label{eq:NLO_mu}
O_s^{(6,1)}(C_i(\mu),\mu) = O_s^{(6,1)}(C_i(\mu),M_Z)
- v_s^2 O_s^{(6,0)}(\dot{C}_i(\mu)) \ln\frac{\mu}{M_Z} \, , 
\end{align}
and we have checked that our results indeed satisfy this equation.

\section{Results and uncertainty estimates}
\label{sec:results}
The main results of this work are analytic expressions for the EWPO listed in Eqs.~\eqref{eq:EWPO_W}-\eqref{eq:EWPO_add}.  Results for the EWPO at LO
in SMEFT in the different EW input schemes can be constructed from partial 
decay widths of the $W$ and $Z$ boson into left- and right-handed fermions, which are
given in Appendix~\ref{sec:EWPO_LO}.  Moreover, in the limit of vanishing quark
masses, the NLO QCD corrections appearing in $W$ and $Z$ decays into quarks
are a universal correction to the LO SMEFT results.  Explicitly,
 \begin{align}
 \label{eq:NLO_QCD}
 \frac{\Gamma^{\rm QCD}(Z\to q\bar{q})}{\Gamma^{\rm LO}(Z\to q\bar{q})} = 
  \frac{\Gamma^{\rm QCD}(W\to ud)}{\Gamma^{\rm LO}(W\to ud)} = \frac{3}{4}\frac{C_F\alpha_s(\mu_R)}{\pi} \,,
 \end{align}
where $C_F=4/3$ in QCD.  This universality implies that NLO QCD corrections cancel out of EWPO based on
ratios of quark decays; the set of such EWPO considered in this work is
$R_b,R_c,A_b,A_c,A_{\rm FB}^b,A_{\rm FB}^c$.

The NLO electroweak corrections are rather lengthy and are not reproduced here.
Instead, the full NLO results (including the QCD corrections just discussed) are provided 
along with the LO ones in ancillary computer files with  the arXiv submission of this work,
and can be evaluated numerically using the included Mathematica notebook.
The notation used for these ancillary files is as follows.  We present each observable at LO and NLO 
in the form
\begin{align}
O_s^{\text{LO}} &= O_s^{(4,0)} + v_s^2 \sum_i \mathtt{Class[i]} \, O_{s,i}^{(6,0)} \, , \nonumber \\
O_s^{\text{NLO}} &= O_s^{(4,0)} + \frac{1}{v_s^2} O_s^{(4,1)} + \sum_i \mathtt{Class[i]} \, \left(v_s^2 O_{s,i}^{(6,0)} + O_{s,i}^{(6,1)} \right) \, , 
\end{align}
where the summation index $i$ runs over the eight operator classes of the Warsaw basis. 
The tags \texttt{Class[i]} (where here and below the font used in ``\texttt{Object}'' refers to a variable or file name in the electronic results) allow to separate contributions from specific 
operator classes or the SM. 
Predictions for the individual observables are given as separate files with the filename \\

\noindent
 \texttt{<observable>\_SMEFT\_<order>\_<scheme>.m}, \\
 
\noindent where \texttt{order} is either \texttt{LO} or \texttt{NLO} and the \texttt{scheme} is referred to by a list of their input parameters, see Tab.~\ref{tab:schemeDef}. 
Note again, that in contrast to~\cite{Biekotter:2023xle,Biekotter:2023vbh}, we present our results for a covariant derivative defined with a plus sign to agree with the \texttt{SMEFTsim} conventions~\cite{Brivio:2020onw}.
We list the variables used in our results in Tab.~\ref{tab:variables_in_results}. 
Wilson coefficient names are written in the conventions of \texttt{DsixTools} with the exception of $C_W$ which we write as \texttt{CWw} to distinguish it from the cosine of the Weinberg angle written as \texttt{CW}.
PV integrals are given in the notation of \texttt{LoopTools}~\cite{Hahn:1998yk}, and only the finite
part is to be used. These PV integrals can also be substituted by standard functions by applying the replacements \texttt{IntRep} in our ancillary notebook. An example is given in the notebook. 

\begin{table}
\centering
\begin{tabular}{lrl}
variable & definition & description \\
\hline
\texttt{MX} & $M_X$ & mass of particle $X$ \\
\texttt{MX2} & $M_X^2$ & square of the mass of particle $X$ \\
\hline
\texttt{SW} & $s_w$ & sine of the Weinberg angle\\
\texttt{CW} & $c_w$ & cosine of the Weinberg angle\\
\texttt{SW2} & $s_w^2$ & sine of the Weinberg angle squared\\
\texttt{CW2} & $c_w^2$ & cosine of the Weinberg angle squared\\
\texttt{CWw} & $C_W$ & Wilson coefficient $C_W$ (to distinguish it from $c_w$)\\
\hline
\texttt{VMU} & $v_\mu$ & vev in schemes using $G_F$ as an input parameter \\
\texttt{VALPHA} & $v_\alpha$ & vev in the $\{\alpha, M_W, M_Z \}$ scheme \\
\texttt{VALPHAEFF} & $v_\alpha^{\rm eff}$ & vev in the $\{\alpha, \seff, M_Z \}$ scheme \\
\hline
\texttt{Class[k]} & tag & tag for SMEFT contributions from class $k$ \\
\texttt{Leff} &  & Flavour index of the reference lepton \\
\hline
\texttt{A0i}, \texttt{B0i}, \texttt{C0i}  & & PV integrals in the notation of \texttt{LoopTools} \\
\hline
\end{tabular}
\caption{Variables used in the ancillary files. $X=Z,W,H,T$ for the $Z$, $W$ and Higgs boson and the top quark, respectively.}
\label{tab:variables_in_results}
\end{table}

Realistic models of new physics typically do not match onto a set of Wilson coefficients with a completely arbitrary flavour structure. 
Instead, potential symmetries of the new physics model will manifest themselves in the structure and correlations of the SMEFT Wilson coefficients. 
The possibility that the SM Yukawa couplings are the only sources of the breaking of the flavour symmetry is known under the name of minimal flavour violation~(MFV)~\cite{Gerard:1982mm,Chivukula:1987py,Hall:1990ac,DAmbrosio:2002vsn}. 
Two lists of replacements in the notebook accompanying this publication allow to write the results in terms of the independent Wilson coefficients under the $U(3)^5$ and MFV assumptions. Details on the notation of the Wilson coefficients under these assumptions are given in appendix~\ref{sec:flav}. 

In any phenomenological study,  it is important to have a means of estimating uncertainties from
uncalculated higher-order corrections in the SMEFT expansion. Within a given electroweak input scheme, a  typical way to estimate higher-order perturbative corrections is to study
the stability of results under variations of the SMEFT renormalisation scale~$\mu$ around a 
particular default value, which for EWPO is typically chosen as $\mu^{\rm def} = M_Z$. 
Taking the Wilson coefficients $C_i(M_Z)$ as unknown parameters, as is the case in a global fit, 
in order to perform $\mu$-variations one uses renormalisation-group (RG) running to express coefficients at arbitrary $\mu$ in terms of those at the default choice $M_Z$.  In the Mathematica notebook, we include the option to vary the scale $\mu$ using the fixed-order  solution of the RG equations, namely
\begin{align}
\label{eq:C_evolve}
C_i(\mu) = C_i(M_Z) +  \dot{C}_i(M_Z) \ln \frac{\mu}{M_Z} \,.
\end{align}
This fixed-order running is sufficient if $\mu\sim M_Z$, which is the case when the 
scale is varied up and down by the customary factors of two. 
More  sophisticated implementations  using the exact solutions to RG equations are also possible but are not considered here. 

The simple form of Eq.~(\ref{eq:C_evolve}) and the fact that results depend only linearly on the Wilson coefficients allows to write explicit analytic expressions which make clear how variations of the SMEFT  renormalisation scale $\mu$ estimate higher-order effects.\footnote{
In the present discussion, we do not vary the
QCD renormalisation scale appearing through $\alpha_s(\mu_R)$ in the NLO QCD corrections Eq.~(\ref{eq:NLO_QCD}); in the Mathematica notebook  correlated variations $\mu_R=\mu$
are implemented, although independent variations of $\mu_R$ and $\mu$ are also possible.}
At LO, one has
\begin{align}
 O_s^{(6,0)}(C_i(\mu) ) -   O_s^{(6,0)}(C_i(M_Z) )=  \ln \frac{\mu}{M_Z} O_s^{(6,0)}(\dot{C}_i(M_Z) ) \, ,
\end{align}
where the term on the right-hand side is of one-loop order and taken as an
indication of beyond-LO corrections. Similarly, at NLO once can use Eqs.~(\ref{eq:Cdot_gamma}, \ref{eq:NLO_mu}) and  Eq.~(\ref{eq:C_evolve}) to arrive at 
\begin{align}
O_s^{\rm NLO}(C_i(\mu), \mu)-O_s^{\rm NLO}(C_i(M_Z), M_Z) = &  \ln \frac{\mu}{M_Z} O_s^{(6,1)}(\dot{C}_i(M_Z),M_Z) \nonumber \\ &
-   \ln^2\left( \frac{\mu}{M_Z}\right) O_s^{(6,0)}(\gamma_{ij}^{(4,1)}\dot{C}_j(M_Z) )\, ,
\end{align}
where in the last term on the right-hand side the vev dependence of the $(6,0)$ contribution has cancelled against the one we have pulled out of the definition of the anomalous dimension matrix, see Eq.~\eqref{eq:Cdot_gamma}. In this case, the two terms on the right-hand side are both 
NNLO in the couplings
and thus give an indication of the size of higher-order corrections to the dimension-6
results. On the other hand, these are both products of one-loop quantities and thus miss genuine two-loop effects in the  SMEFT anomalous dimension, which are currently unknown.  

Given the somewhat arbitrary nature of uncertainty estimates based on scale variations, and also
the fact that apart from $\alpha_s$ the SM parameters appearing in EWPO are renormalised
on-shell and thus contain no scale to vary, it is important to have additional means of estimating
uncertainties.  In both the SM and SMEFT, a simple way to do this is to calculate observables 
in several different electroweak input schemes, which are equivalent at a given order in the SMEFT expansion, but organise higher-order corrections in both operator-dimension and loops differently.  It 
is precisely for this reason that we have given results in five different input schemes, and advocate
their use in global SMEFT fits.

\FloatBarrier
\section{Conclusions}
\label{sec:conclusions}
We have presented analytic results for EWPO to NLO in dimension-six SMEFT in the five different EW input schemes listed in Table~\ref{tab:schemeDef}.   These results will be useful for SMEFT analyses of data from current and future lepton colliders, which play an important role in global fits. 
A Mathematica notebook for the numerical evaluation of the results is provided with the arXiv submission of the paper, in which numerical inputs can be adjusted as needed to facilitate the combination of EWPO analyses with other observables. While our results have made no assumptions on the flavour structure of SMEFT Wilson coefficients, we have included in the notebook options to implement $U(3)^5$ and MFV assumptions.   Furthermore, theory uncertainties for specific 
observables or in global fits can be  estimated through scale variations, calculating in different
EW input schemes, or preferably both.  

\section*{Acknowledgements}
We thank Tommy Smith for valuable discussions and for his contributions during the early stages of this work. 
This research was supported by the Deutsche Forschungsgemeinschaft (DFG, German Research Foundation) under grant 396021762 - TRR 257.

 \appendix

\section{EWPO on the $Z$ pole}
\label{sec:EWPO_def}

In this appendix, we define the EWPO on the $Z$ pole in terms of the $Z$-boson partial decay rates. 
The ratios $R_x$ are defined as
\begin{align}
R_\ell &= \frac{\sum_q \Gamma (Z \to q \bar{q})}{\Gamma (Z \to \ell \ell)} \, ,
\nonumber \\
R_{q_i} &= \frac{\Gamma (Z \to q_i \bar{q}_i) }{ \sum_q \Gamma (Z \to q \bar{q})} \, , 
\end{align}
with $\ell=e, \mu, \tau$ and $q_i=s, c, b$ and the sum runs over $q=u,d,c,s,b$. 
The left-right asymmetries $A_x$ are defined as 
\begin{align}
A_x &= \frac{\Gamma(Z \to x_L \bar{x}_L) - \Gamma(Z \to x_R \bar{x}_R)}{\Gamma(Z \to x \bar{x})} \, .
\end{align}
The asymmetry $A_\ell$ is directly related to the effective weak mixing angle $\seff$ (using the same reference lepton $\ell$) defined as 
\begin{align}
\label{eq:seff_Aell}
\seff & = \frac{1}{2} \left( \frac{\Gamma(Z \to \ell_L \bar{\ell}_L)}{\Gamma(Z \to \ell_L \bar{\ell}_L ) - \Gamma(Z \to \ell_R \bar{\ell}_R)} \right) 
= \frac{1}{4} \left( 1 - \frac{1}{A_\ell} \right) \, .
\end{align}
Therefore, only one of the quantities $A_\ell$ and $\seff$ is an independent quantity. 
The forward-backward asymmetries $\AFB^x$ are defined as 
\begin{align}
\AFB^x &= \frac{\sigma_F - \sigma_B}{\sigma_F + \sigma_B} = \frac{3}{4} A_x A_\ell \, ,
\end{align}
where $\sigma_F$ and $\sigma_B$ are defined by the angle $\theta$ between the incoming lepton $\ell^-$ and the outgoing anti-fermion $\bar{f}$ being within $\theta \in [0, \, \pi/2]$ and $\theta \in [\pi/2, \, \pi]$, respectively. 
The hadronic total cross section for the process $e^+ e^- \to \text{hadrons}$ can be parametrised in the narrow width approximation as 
\begin{align}
\sigma_{\text{had}} &= \sum_{q = u, d, c, s, b} \frac{12 \pi}{M_Z} \frac{\Gamma_e \Gamma_q}{\Gamma_Z^2} \, .
\end{align}

\section{Flavour assumptions}
\label{sec:flav}
%
To facilitate the interpretation of our results under the assumption of common flavour assumptions, 
the replacement lists \texttt{flavU35} and \texttt{flavMFV} in our ancillary notebook allow to express the results under the $U(3)^5$ and MFV assumption. 
A nice summary of the relevant simplifications, based on Refs.~\cite{Faroughy:2020ina,Bruggisser:2021duo},  is given in Sections II~B and II~C of~\cite{Bellafronte:2023amz}. Instead of reproducing the necessary equations here, we only clarify the notation in which we present the Wilson coefficients. 

\subsection{$U(3)^5$ symmetry}
A rather strict requirement on the Wilson coefficients of the SMEFT is the assumption of a $U(3)$ symmetry for all fermion fields
\begin{equation}
U(3)^5=U(3)_\ell \times U(3)_q\times U(3)_e\times U(3)_u\times U(3)_d,
\end{equation}
where $\{\ell, q, e, u, d\}$ represent the SM fermions. 
Under this assumption, there are 41+6 independent CP-even+CP-odd SMEFT operators~\cite{Faroughy:2020ina}. 
Operators with two flavour indices have a single independent Wilson coefficient and we hence drop the flavour indices for these coefficients. 
Note that dipole operators like $Q_{uW}$ and $Q_{uB}$, which generally contribute to EWPO are completely forbidden under a $U(3)^5$ symmetry. 
Four-fermion operators with two different fermion bilinears again have a single independent Wilson coefficient, for which we thus also drop the flavour indices.  

Four-fermion operator structures with two fermion currents of the same chirality, explicitly $Q_{qq}^{(1)}$, $Q_{qq}^{(3)}$, $Q_{ll}$, $Q_{dd}$ and $Q_{uu}$, generally have two independent $U(3)^5$ singlets. 
We refer to these two independent structures with unprimed and primed Wilson coefficients, corresponding to the operators contracting the flavour indices within the fermion bilinears or between the two different fermion bilinears, respectively.  
As an example, the operator $Q_{\substack{ll\\ijkl}} = (\bar l_i \gamma_\mu l_j)(\bar l_k \gamma^\mu l_l)$ has two flavour-symmetric contractions 
\begin{align}
C_{ll} \, \delta_{ij}\delta_{lk}  \qquad \text{and} \qquad  C_{ll}^{\prime} \, \delta_{ik}\delta_{jl} \, .
\end{align}
In our numerical results, we refer to the primed coefficients by adding the letter \texttt{p} to the Wilson coefficient names, for instance we replace the different flavour combinations of $C_{ll}$ as 
\begin{align}
C_{\substack{ll\\1122}} & \to \mathtt{Cll} \, , 
& 
C_{\substack{ll\\1221}} & \to \mathtt{Cllp} \, .
\end{align}
For the operator $Q_{ee}$, a Fierz identity implies a single independent coefficient $C_{ee}$. 

\subsection{Minimal flavour violation}
A more general flavour symmetry for the Wilson coefficients is given by MFV, which assumes the SM Yukawa couplings are the only sources of the breaking of the $U(3)^5$ flavour symmetry. 
MFV is thus an expansion in powers of the Yukawa couplings. 
As we are assuming that all fermions except the top are massless, right-handed down-type quarks still retain a $U(3)$ symmetry under our implementation of MFV.
Wilson coefficients involving left-handed third-generation quark couplings or right-handed up-type quark couplings are independent of those of the first/second generation. 
Instead of keeping the flavour indices, we use the subscripts (\texttt{Q}, \texttt{t}) for the third generation and (\texttt{q},  \texttt{u}) for the first/second generation in the Wilson coefficients. 
For Wilson coefficients with two indices, we write, for instance
\begin{align}
C_{\substack{Hq\\22}}^{(3)} & \to \mathtt{CHq3} \, , 
& 
C_{\substack{Hq\\33}}^{(3)} & \to \mathtt{CHQ3} \, .
\end{align}
The four-fermion Wilson coefficients with two equal fermion bilinears simplify as 
\begin{align}
C_{\substack{uu\\1122}} & \to \mathtt{Cuu} \, , 
& 
C_{\substack{uu\\1133}} & \to \mathtt{Cut} \, ,
& 
C_{\substack{uu\\1221}} & \to \mathtt{Cuup} 
\nonumber \\
C_{\substack{uu\\1331}} & \to \mathtt{Cutp} \, ,
&
C_{\substack{uu\\3333}} & \to 2 \left( \mathtt{Cut} +\mathtt{Cutp} \right) - \mathtt{Cuu} - \mathtt{Cuup}
\, .
\end{align}
There are four independent coefficients for operators with four up-type quarks, and two coefficients for those with two up-type quarks. 
Other four-fermion operators simplify in the same way as under a $U(3)^5$ symmetry.

\section{Comparison with previous work}
\label{sec:comparison_prev}

Numerical results for EWPO in an input-scheme involving $\{G_F , \alpha , M_Z \}$
have been previously published in~\cite{Dawson:2019clf,Dawson:2022bxd,Bellafronte:2023amz}. 
These are closely related to our results in the LEP scheme, but differ in the following ways:
\begin{enumerate}
	\item[a)] Definition of the electromagnetic coupling constant $\alpha$: 
		While we use the on-shell value at the $Z$-boson mass, $\alpha(M_Z)$, as an input, \cite{Dawson:2019clf,Dawson:2022bxd,Bellafronte:2023amz} employs $\alpha(0)$. 
		The two choices are related by Eq.~\eqref{eq:alpha_relations} and the light fermion contributions to $\alpha$ are included in the NLO expansion coefficients in  \cite{Dawson:2019clf,Dawson:2022bxd,Bellafronte:2023amz}, rather than being absorbed into the definition of $\alpha(M_Z)$ as in our LEP scheme.
	\item[b)] Partial higher-order corrections are included in~\cite{Dawson:2019clf,Dawson:2022bxd,Bellafronte:2023amz}, see Eq.~(58) of~\cite{Bellafronte:2023amz}, through the numerical evaluation of the $W$-boson mass at its best theory prediction in LO contributions and at its NLO value in NLO contributions to observables. 
	\item[c)] Due to a different sign convention in the covariant derivative, our predictions have opposite signs for operators including an odd number of field strengths, specifically $C_W$, $C_{uB}$ and $C_{uW}$. 
\end{enumerate}

Comparing numerically, we exactly agree on all LO predictions after switching to the numerical values used in~\cite{Dawson:2019clf,Dawson:2022bxd,Bellafronte:2023amz} (including replacing $\alpha(M_Z)$ by $\alpha(0)$).
At NLO, we find good numerical agreement for the decay rates $\Gamma_W$ and $\Gamma_Z$ despite the different choices for $\alpha$ and differences in the expansions. 
For observables based on ratios or differences of two similar-size contributions, individual Wilson coefficients (appearing first at NLO) experience larger differences.

\section{Analytic results for EWPO at LO in SMEFT}
\label{sec:EWPO_LO}

The EWPO in Eqs.~(\ref{eq:EWPO_W}, \ref{eq:EWPO_Z})
can be derived from $W$ and $Z$-boson decays into left- and right-handed
fermions.  In what follows, we give compact results for such partial decay widths in the five EW input schemes used in this work at LO in SMEFT.  These expressions involve products of functions which
receive both SM and dimension-6 contributions; it is understood that to obtain the decay rates
and consequently EWPO one must expand out the products and retain only up to linear corrections in the SMEFT  Wilson coefficients.  Results for the additional observables Eq.~(\ref{eq:EWPO_add})
are given at the end of the section.

For $Z$ decay, we can write
\begin{align}
\Gamma(Z\to f\bar{f})& = M_Z \frac{N_c^f}{24\pi}|{\cal N}_Z|^2\left(|{\cal Z}^f_{L}|^2 +|{\cal Z}^f_R|^2 \right)\,,
\end{align}
where $N_c^f$ was defined after Eq.~(\ref{eq:Alpha_Convert}). The normalisation factor is 
\begin{align}
\label{eq:NZ}
{\cal N}_Z = \frac{M_Z}{v_s}\left[1- v_s^2 \left( \frac{1}{4} C_{HD}+\frac{1}{2}\Delta v_s^{(6,0)}\right)   \right]\, ,
\end{align}
where the $\Delta v_s$ depend on the scheme $s$ and are listed in Eq.~(\ref{eq:DV60})
below.  We define SMEFT
expansion coefficients for decay into left-handed fermions as 
\begin{align}
{\cal Z}_L^f = {\cal Z}_L^{f(4,0)}+ v_s^2   {\cal Z}_L^{f(6,0)}
\end{align}
and similarly for right-handed decays. In the SM, one has 
\begin{align}
 {\cal Z}_L^{f(4,0)}  = 2 T_3^f - 2 s_w^2 Q_f \, , \qquad  {\cal Z}_R^{f(4,0)}  =   - 2 s_w^2 Q_f \,,
\end{align}
where $Q_f$ is the charge and $T_3^f$ is the third component of the  weak isospin of fermion $f$.\footnote{Our
convention is such that $Q_u = 2/3, Q_e = -1$, while $T_3^u=1/2$, and so on. Here and below the definitions of $M_W$ and $s_w$ depend on the scheme $s$ and 
should be understood as written in Eq.~(\ref{eq:CW_def}).  }
 The result in SMEFT can be written as 
\begin{align}
 {\cal Z}_{L/R}^{f(6,0)} & =Q_f \left(-G^{(6,0)} + 4  c_w^2  \Delta_W^s\right)+  g^{f(6,0)}_{L/R}  \,.
\end{align}
The above results contain, first off, the fermion-species independent function 
\begin{align}
	\label{eq:Gdef}
	G^{(6,0)} & =  -c_w^2 C_{HD} - 2 c_w s_w C_{HWB}  \, , 
	\end{align}
and secondly fermion-specific functions, which for charged leptons $\ell_i$ and neutrinos $\nu_i$ are
 \begin{align}
	\label{eq:gl}
 g^{\ell(6,0)}_L & = -   C_{\substack{Hl \\ ii}}^{(1)}  - C_{\substack{Hl\\ ii}}^{(3)}\, , \qquad
g^{\ell(6,0)}_R  = -   C_{\substack{He\\ ii}} \,, \nonumber \\
 g^{\nu(6,0)}_L & = -   C_{\substack{Hl \\ ii}}^{(1)}  + C_{\substack{Hl\\ ii}}^{(3)}\, , \qquad
g^{\nu(6,0)}_R  =0 \, ,
\end{align} 
while for up and down-type quarks
\begin{align}
	\label{eq:gq}
	g^{d(6,0)}_L & = -   C_{\substack{Hq \\ ii}}^{(1)}  - C_{\substack{Hq\\ ii}}^{(3)}\, , \qquad
g^{d(6,0)}_R  =  - C_{\substack{Hd\\ ii}} \,, \nonumber \\
 g^{u(6,0)}_L & = -   C_{\substack{Hq \\ ii}}^{(1)}  + C_{\substack{Hq\\ ii}}^{(3)}\, , \qquad
g^{u(6,0)}_R  = -   C_{\substack{Hu\\ ii}} \,.
\end{align} 
Finally, in schemes $s$ where $M_W$ is not an input, the quantity $\Delta_W^s$ appears 
in the following LO SMEFT relation between the on-shell and derived masses:
\begin{align}
\label{eq:MW_LO}
M_W^{\rm LO} = M_Z c_w\left(1 + v_s^2 \Delta_W^{s(6,0)}\right) \,; \qquad s\in \{v_\alpha^{\rm eff},v_\mu^{\rm eff}, {\rm LEP}\}\,.
\end{align}
The explicit results are 
\begin{align}
\Delta_W^{{\rm LEP}(6,0)}&  = -\frac{s_w^2}{2c_{2w}}\left[\Delta v_\mu^{(6,0)} - \Delta v_\alpha^{(6,0)} \right]  \, , \\
\Delta_W^{ v_\sigma^{\rm eff}(6,0)}&= \frac{1}{4c_w^2}\left[G^{(6,0)} + 2s_w^2 g_L^{\ell(6,0)}
+ c_{2w}g_R^{\ell(6,0)} \right]       \,,
\end{align}
where $\sigma\in\{\alpha,\mu\}$ and $c_{2w}\equiv 1-2s_w^2$.
 Here and in Eq.~(\ref{eq:NZ}), the tree-level vev shifts are
\begin{align}
\label{eq:DV60}
\Delta v_\mu^{(6,0)} & =  C_{\substack{Hl \\ 11}}^{(3)} + C_{\substack{Hl \\ 22}}^{(3)}
  - C_{\substack{ll \\ 1221}}  \, , \nonumber \\
  \Delta v_\alpha^{(6,0)} & =   -2 \frac{c_w}{s_w}
\left[C_{HWB} + \frac{c_w}{4 s_w}C_{HD} \right] \, , \nonumber \\
\Delta v_\alpha^{\rm eff} & =   -\frac{1}{2}C_{HD} - \frac{1}{c_w s_w}C_{HWB}- \frac{c_{2w}}{c_w^2}\left(g_L^{\ell(6,0)} + \frac{c_{2w}}{2s_w^2}g_R^{\ell(6,0)} \right) \, ,
\end{align}
where one is to use $\Delta v_\mu$ for $s\in\{\alpha_\mu,v_\mu^{\rm eff}, {\rm LEP}\}$, while for
$s=\alpha$ or $s=v_\alpha^{\rm eff}$ one uses $\Delta v_\alpha$ or $\Delta v_\alpha^{\rm eff}$,
respectively.
 
 The results for $W\to f f'$ are considerably more compact. In this case we can write
 \begin{align}
\Gamma(W\to ff')& = M_W\left(1 + v_s^2 \Delta_W^{s(6,0)}\right) \frac{N_c^f}{24\pi} \left|{\cal N}_W\right|^2\left|1 + v_s^2 \, {\cal W}^{ff'(6,0)}\right|^2 \,,
\end{align}
where the normalisation factor is
 \begin{align}
 {\cal N}_W = \frac{\sqrt{2}M_W}{v_s}\left[1 + v_s^2\left(\Delta_W^{s(6,0)} -\frac{1}{2}\Delta v_s^{(6,0)} \right)\right] \,,
 \end{align}
and the fermion-specific functions  for leptonic and hadronic decays are given by
 \begin{align}
 {\cal W}^{\ell\nu(6,0)} & = C_{\substack{Hl \\ ii}}^{(3)} \, , \qquad 
  {\cal W}^{ud(6,0)}  = C_{\substack{Hq \\ ii}}^{(3)}  \, .
  \end{align}
 
 In schemes where they are not inputs, the predictions for $M_W, G_F$ and $\alpha(M_Z)$ 
 provide additional EWPO.  LO results for the on-shell $W$ boson mass in such
 schemes have already been given in Eq.~(\ref{eq:MW_LO}).  For $G_F$, one has 
 \begin{align}
G_F^{\rm LO} & = \frac{1}{\sqrt{2} v_s^2}\left[1 - v_s^2\left( \Delta v_s^{(6,0)} - \Delta v_\mu^{(6,0)} \right)\right]\, ;  \qquad s\in \{\alpha,v_\alpha^{\rm eff}\} \, ,
 \end{align}
while for $\alpha(M_Z)$, one has instead
 \begin{align}
\alpha(M_Z)^{\rm LO} & = \frac{M_Z^2 c_w^2 s_w^2}{\pi v_\mu^2}\left[1 + v_\mu^2\left(\Delta v_\mu^{(6,0)} - \Delta v_{\bar{s}}^{(6,0)} \right)\right]\, ;  \quad \bar{s}\in \{\alpha \text{ for } s=\alpha_\mu ,v_\alpha^{\rm eff} \text{ for }s=v_\mu^{\rm eff}\}\,,
 \end{align}
where $\bar{s}$ refers to the scheme which shares two inputs with $s$, but differs by using $\alpha(M_Z)$ instead of $G_F$.

\newpage
\begin{table}
\begin{center}
\small
\begin{minipage}[t]{4.4cm}
\renewcommand{\arraystretch}{1.5}
\begin{tabular}[t]{c|c}
\multicolumn{2}{c}{$1:X^3$} \\
\hline
$Q_G$                & $f^{ABC} G_\mu^{A\nu} G_\nu^{B\rho} G_\rho^{C\mu} $ \\
$Q_{\widetilde G}$          & $f^{ABC} \widetilde G_\mu^{A\nu} G_\nu^{B\rho} G_\rho^{C\mu} $ \\
$Q_W$                & $\epsilon^{IJK} W_\mu^{I\nu} W_\nu^{J\rho} W_\rho^{K\mu}$ \\ 
$Q_{\widetilde W}$          & $\epsilon^{IJK} \widetilde W_\mu^{I\nu} W_\nu^{J\rho} W_\rho^{K\mu}$ \\
\end{tabular}
\end{minipage}
%
\begin{minipage}[t]{2.5cm}
\renewcommand{\arraystretch}{1.5}
\begin{tabular}[t]{c|c}
\multicolumn{2}{c}{$2:H^6$} \\
\hline
$Q_H$       & $(H^\dag H)^3$ 
\end{tabular}
\end{minipage}
\begin{minipage}[t]{4.9cm}
\renewcommand{\arraystretch}{1.5}
\begin{tabular}[t]{c|c}
\multicolumn{2}{c}{$3:H^4 D^2$} \\
\hline
$Q_{H\Box}$ & $(H^\dag H)\Box(H^\dag H)$ \\
$Q_{H D}$   & $\ \left(H^\dag D_\mu H\right)^* \left(H^\dag D_\mu H\right)$ 
\end{tabular}
\end{minipage}
%
\begin{minipage}[t]{2.5cm}
\renewcommand{\arraystretch}{1.5}
\begin{tabular}[t]{c|c}
\multicolumn{2}{c}{$5: \psi^2H^3 + \hbox{h.c.}$} \\
\hline
$Q_{eH}$           & $(H^\dag H)(\bar l_p e_r H)$ \\
$Q_{uH}$          & $(H^\dag H)(\bar q_p u_r \widetilde H )$ \\
$Q_{dH}$           & $(H^\dag H)(\bar q_p d_r H)$\\
\end{tabular}
\end{minipage}

\begin{minipage}[t]{4.7cm}
\renewcommand{\arraystretch}{1.5}
\begin{tabular}[t]{c|c}
\multicolumn{2}{c}{$4:X^2H^2$} \\
\hline
$Q_{H G}$     & $H^\dag H\, G^A_{\mu\nu} G^{A\mu\nu}$ \\
$Q_{H\widetilde G}$         & $H^\dag H\, \widetilde G^A_{\mu\nu} G^{A\mu\nu}$ \\
$Q_{H W}$     & $H^\dag H\, W^I_{\mu\nu} W^{I\mu\nu}$ \\
$Q_{H\widetilde W}$         & $H^\dag H\, \widetilde W^I_{\mu\nu} W^{I\mu\nu}$ \\
$Q_{H B}$     & $ H^\dag H\, B_{\mu\nu} B^{\mu\nu}$ \\
$Q_{H\widetilde B}$         & $H^\dag H\, \widetilde B_{\mu\nu} B^{\mu\nu}$ \\
$Q_{H WB}$     & $ H^\dag \sigma^I H\, W^I_{\mu\nu} B^{\mu\nu}$ \\
$Q_{H\widetilde W B}$         & $H^\dag \sigma^I H\, \widetilde W^I_{\mu\nu} B^{\mu\nu}$ 
\end{tabular}
\end{minipage}
%
\begin{minipage}[t]{5.2cm}
\renewcommand{\arraystretch}{1.5}
\begin{tabular}[t]{c|c}
\multicolumn{2}{c}{$6:\psi^2 XH+\hbox{h.c.}$} \\
\hline
$Q_{eW}$      & $(\bar l_p \sigma^{\mu\nu} e_r) \sigma^I H W_{\mu\nu}^I$ \\
$Q_{eB}$        & $(\bar l_p \sigma^{\mu\nu} e_r) H B_{\mu\nu}$ \\
$Q_{uG}$        & $(\bar q_p \sigma^{\mu\nu} T^A u_r) \widetilde H \, G_{\mu\nu}^A$ \\
$Q_{uW}$        & $(\bar q_p \sigma^{\mu\nu} u_r) \sigma^I \widetilde H \, W_{\mu\nu}^I$ \\
$Q_{uB}$        & $(\bar q_p \sigma^{\mu\nu} u_r) \widetilde H \, B_{\mu\nu}$ \\
$Q_{dG}$        & $(\bar q_p \sigma^{\mu\nu} T^A d_r) H\, G_{\mu\nu}^A$ \\
$Q_{dW}$         & $(\bar q_p \sigma^{\mu\nu} d_r) \sigma^I H\, W_{\mu\nu}^I$ \\
$Q_{dB}$        & $(\bar q_p \sigma^{\mu\nu} d_r) H\, B_{\mu\nu}$ 
\end{tabular}
\end{minipage}
%
\begin{minipage}[t]{5cm}
\renewcommand{\arraystretch}{1.5}
\begin{tabular}[t]{c|c}
\multicolumn{2}{c}{$7:\psi^2H^2 D$} \\
\hline
$Q_{H l}^{(1)}$      & $(H^\dag i\overleftrightarrow{D}_\mu H)(\bar l_p \gamma^\mu l_r)$\\
$Q_{H l}^{(3)}$      & $(H^\dag i\overleftrightarrow{D}^I_\mu H)(\bar l_p \sigma^I \gamma^\mu l_r)$\\
$Q_{H e}$            & $(H^\dag i\overleftrightarrow{D}_\mu H)(\bar e_p \gamma^\mu e_r)$\\
$Q_{H q}^{(1)}$      & $(H^\dag i\overleftrightarrow{D}_\mu H)(\bar q_p \gamma^\mu q_r)$\\
$Q_{H q}^{(3)}$      & $(H^\dag i\overleftrightarrow{D}^I_\mu H)(\bar q_p \sigma^I \gamma^\mu q_r)$\\
$Q_{H u}$            & $(H^\dag i\overleftrightarrow{D}_\mu H)(\bar u_p \gamma^\mu u_r)$\\
$Q_{H d}$            & $(H^\dag i\overleftrightarrow{D}_\mu H)(\bar d_p \gamma^\mu d_r)$\\
$Q_{H u d}$ + h.c.   & $i(\widetilde H ^\dag D_\mu H)(\bar u_p \gamma^\mu d_r)$\\
\end{tabular}
\end{minipage}

\vspace{0.25cm}

\begin{minipage}[t]{4.75cm}
\renewcommand{\arraystretch}{1.5}
\begin{tabular}[t]{c|c}
\multicolumn{2}{c}{$8:(\bar LL)(\bar LL)$} \\
\hline
$Q_{ll}$        & $(\bar l_p \gamma_\mu l_r)(\bar l_s \gamma^\mu l_t)$ \\
$Q_{qq}^{(1)}$  & $(\bar q_p \gamma_\mu q_r)(\bar q_s \gamma^\mu q_t)$ \\
$Q_{qq}^{(3)}$  & $(\bar q_p \gamma_\mu \sigma^I q_r)(\bar q_s \gamma^\mu \sigma^I q_t)$ \\
$Q_{lq}^{(1)}$                & $(\bar l_p \gamma_\mu l_r)(\bar q_s \gamma^\mu q_t)$ \\
$Q_{lq}^{(3)}$                & $(\bar l_p \gamma_\mu \sigma^I l_r)(\bar q_s \gamma^\mu \sigma^I q_t)$ 
\end{tabular}
\end{minipage}
\begin{minipage}[t]{5.25cm}
\renewcommand{\arraystretch}{1.5}
\begin{tabular}[t]{c|c}
\multicolumn{2}{c}{$8:(\bar RR)(\bar RR)$} \\
\hline
$Q_{ee}$               & $(\bar e_p \gamma_\mu e_r)(\bar e_s \gamma^\mu e_t)$ \\
$Q_{uu}$        & $(\bar u_p \gamma_\mu u_r)(\bar u_s \gamma^\mu u_t)$ \\
$Q_{dd}$        & $(\bar d_p \gamma_\mu d_r)(\bar d_s \gamma^\mu d_t)$ \\
$Q_{eu}$                      & $(\bar e_p \gamma_\mu e_r)(\bar u_s \gamma^\mu u_t)$ \\
$Q_{ed}$                      & $(\bar e_p \gamma_\mu e_r)(\bar d_s\gamma^\mu d_t)$ \\
$Q_{ud}^{(1)}$                & $(\bar u_p \gamma_\mu u_r)(\bar d_s \gamma^\mu d_t)$ \\
$Q_{ud}^{(8)}$                & $(\bar u_p \gamma_\mu T^A u_r)(\bar d_s \gamma^\mu T^A d_t)$ \\
\end{tabular}
\end{minipage}
\begin{minipage}[t]{4.75cm}
\renewcommand{\arraystretch}{1.5}
\begin{tabular}[t]{c|c}
\multicolumn{2}{c}{$8:(\bar LL)(\bar RR)$} \\
\hline
$Q_{le}$               & $(\bar l_p \gamma_\mu l_r)(\bar e_s \gamma^\mu e_t)$ \\
$Q_{lu}$               & $(\bar l_p \gamma_\mu l_r)(\bar u_s \gamma^\mu u_t)$ \\
$Q_{ld}$               & $(\bar l_p \gamma_\mu l_r)(\bar d_s \gamma^\mu d_t)$ \\
$Q_{qe}$               & $(\bar q_p \gamma_\mu q_r)(\bar e_s \gamma^\mu e_t)$ \\
$Q_{qu}^{(1)}$         & $(\bar q_p \gamma_\mu q_r)(\bar u_s \gamma^\mu u_t)$ \\ 
$Q_{qu}^{(8)}$         & $(\bar q_p \gamma_\mu T^A q_r)(\bar u_s \gamma^\mu T^A u_t)$ \\ 
$Q_{qd}^{(1)}$ & $(\bar q_p \gamma_\mu q_r)(\bar d_s \gamma^\mu d_t)$ \\
$Q_{qd}^{(8)}$ & $(\bar q_p \gamma_\mu T^A q_r)(\bar d_s \gamma^\mu T^A d_t)$\\
\end{tabular}
\end{minipage}

\vspace{0.25cm}

\begin{minipage}[t]{3.75cm}
\renewcommand{\arraystretch}{1.5}
\begin{tabular}[t]{c|c}
\multicolumn{2}{c}{$8:(\bar LR)(\bar RL)+\hbox{h.c.}$} \\
\hline
$Q_{ledq}$ & $(\bar l_p^j e_r)(\bar d_s q_{tj})$ 
\end{tabular}
\end{minipage}
\begin{minipage}[t]{5.5cm}
\renewcommand{\arraystretch}{1.5}
\begin{tabular}[t]{c|c}
\multicolumn{2}{c}{$8:(\bar LR)(\bar L R)+\hbox{h.c.}$} \\
\hline
$Q_{quqd}^{(1)}$ & $(\bar q_p^j u_r) \epsilon_{jk} (\bar q_s^k d_t)$ \\
$Q_{quqd}^{(8)}$ & $(\bar q_p^j T^A u_r) \epsilon_{jk} (\bar q_s^k T^A d_t)$ \\
$Q_{lequ}^{(1)}$ & $(\bar l_p^j e_r) \epsilon_{jk} (\bar q_s^k u_t)$ \\
$Q_{lequ}^{(3)}$ & $(\bar l_p^j \sigma_{\mu\nu} e_r) \epsilon_{jk} (\bar q_s^k \sigma^{\mu\nu} u_t)$
\end{tabular}
\end{minipage}
\end{center}
\caption{\label{op59}
The 59 independent baryon number conserving dimension-six operators built from Standard Model fields, in 
the notation of \cite{Jenkins:2013zja}.  The subscripts $p,r,s,t$ are flavour indices, and $\sigma^I$ are Pauli
matrices.}
\end{table}
\FloatBarrier

\bibliography{literature}
\bibliographystyle{JHEP.bst}

\end{document}